# Corrosion-resistant aluminum alloy design through machine learning combined with high-throughput calculations

*Yucheng Ji, Xiaoqian Fu, Feng Ding, Yongtao Xu, Yang He, Min Ao, Fulai Xiao, Dihao Chen, Poulumi Dey, Kui Xiao, Jingli Ren, Xiaogang Li, Chaofang Dong*[*]

Y. Ji, X. Fu, F. Ding, Prof. Y. He, M. Ao, D. Chen, Prof. K. Xiao, Prof. X. Li, and Prof. C. Dong
Beijing Advanced Innovation Center for Materials Genome Engineering
National Materials Corrosion and Protection Data Center
Institute for Advanced Materials and Technology
Shunde Innovation School
University of Science and Technology Beijing, Beijing 100083, China
E-mail: cfdong@ustb.edu.cn

Y. Ji, Dr. P. Dey
Department of Materials Science and Engineering
Faculty of Mechanical, Maritime and Materials Engineering,
Delft University of Technology, Delft 2628CD, The Netherlands

Y. Xu
National Engineering & Technology Research Center for Non-Ferrous Metals Composites
GRINM Group Corporation Limited, Beijing 101407, China

F. Xiao
Shandong Nanshan Aluminum Co., Ltd., Longkou, 265706, China

Prof. J. Ren
Henan Academy of Big Data,
Zhengzhou University, Zhengzhou 450001, China

**Abstract**

Efficiently designing lightweight alloys with combined high corrosion resistance and mechanical properties remains an enduring topic in materials engineering. To this end, machine learning (ML) coupled ab-initio calculations is proposed within this study. Due to the inadequate accuracy of conventional stress-strain ML models caused by corrosion factors, a novel reinforcement self-learning ML algorithm (accuracy $R^2 > 0.92$) is developed. Then, a strategy that integrates ML models, calculated energetics and mechanical moduli is implemented to optimize the Al alloys. Next, this Computation Designed Corrosion-Resistant Al alloy is fabricated that verified the simulation. The performance (elongation reaches ~30%) is attributed to the H-captured Al-Sc-Cu phases (-1.44 eV H$^{-1}$) and Cu-modified η/η' precipitation inside the grain boundaries (GBs). The developed Al-Mg-Zn-Cu interatomic potential (energy accuracy 6.50 meV atom$^{-1}$) proves the cracking resistance of the GB region enhanced by Cu-modification. Conceptually, our strategy is of practical importance for designing new alloys exhibiting corrosion resistance and mechanical properties.

**Significance statement**

For engineering structural component, enhanced mechanical properties facilitate the reduction of material consumption, and excellent corrosion resistance prolongs their service time. Simultaneously improving the mechanical properties and corrosion resistance of materials is a matter of great concern. We present an efficient designing strategy to fabricate light-weight alloys with stronger mechanical properties and corrosion resistance in this paper. We demonstrate our machine learning model combined with high-throughput DFT calculations which discover new structures exhibit superb and comprehensive performance.

**Introduction**

Al alloys, known for their lightweight properties, are widely used in aerospace[1], rail transportation[2], and electric vehicles[3] to reduce energy consumption. As the strongest 7xxx series Al alloys (AA7xxx), its strength is attributed to the dispersion of η/η' phase (Mg-Zn)[4, 5]. However, the large η/η' phase precipitates at the grain boundaries (GBs) weaken the strength of these regions[6]. When AA7xxx are utilized as structural components, they are subjected to environmental H diffusion and accumulation, resulting in stress corrosion cracking (SCC)[7]. Computations revealed that the η/η' phases at the GBs promote H aggregation, reducing the GB cohesive energy by 86.6%[8]. Meanwhile, intergranular cracks (IGCs) are also common SCC morphologies in high-strength Al alloys[9]. Until now, the ultimate tensile strength (UTS) and elongation of anti-SCC AA7xxx reached 672 MPa and 5.02%[10]. Despite the high strength of this alloy makes it more suitable for lightweight applications, this extremely low plasticity poses a significant risk to engineering safety.

To enhance the SCC resistance of Al alloys, one approach is to modify the η/η' phases or retard their consecutive precipitation at the GBs. For instance, switching the η phases to the T phase ($Al_2Mg_3Zn_3$) by heat treatment resulted in a 60% reduction in the cracking areal fractions[11]. Furthermore, crossover Al alloys harmonized by Zn-Mg-Cu have the potential to exhibit excellent performance in mechanical strength and corrosion, while the specific strategies are not stated[12]. High-throughput density function theory (DFT) computations provide wider chemical space for alloy design, the diffusion and stability of 86 kinds of elements in/on the Al matrix/surfaces are confirmed[13]. These calculations identified elements that can diffuse more easily than Zn or Mg thereby avoiding the formation of η/η' phase at the GBs or enhancing the cohesive energy of the GBs, such as Cu[14] and Er[15]. Nevertheless, this qualitative research only provides information about the type of beneficial elements, but the specific composition and heat treatment cannot be determined from such a study. Besides, it is also difficult to determine the reasonable range of multiple beneficial elements in a short period of time only relying on traditional methods.

Machine learning (ML) and big data may be the optimal solutions to this dilemma since they have facilitated research in the field of materials science[16]. However, the biggest problem for metal/alloy prediction is the data scale. Unlike the massive samples of image recognition or real-time speech translation[17], the alloy mechanics (corrosion) dataset size is usually limited to less than 1,000 points[18, 19]. To resolve this, one solution is to reuse the data or specify data labels[20]. Another option is to augment the dataset based on published work or change the predictive descriptor to more accessible features, such as hardness instead of strength[21]. In the case of high-entropy alloys, the dataset can be expanded to 5,000 – 10,000 points using molecular dynamics (MD) simulations[22]. Exploiting calculations to extend ML datasets is a remarkable idea which can achieve the goal of reliable prediction.

For the practical infinite chemical space for alloy design, artificial intelligence combined with physical laws brings us new opportunities[23, 24]. By developing a natural language processing with deep learning, the key descriptors related to pitting potential which can be used to design corrosion-resistant alloys are confirmed[25]. Furthermore, based on random forest algorithms and phases DFT calculations, the corrosion rates of Al alloys under different environmental conditions are predicted[26]. For normal strained strength threshold, the Al fabricated via multiplying η' phases reaches ~800 MPa, while its elongation is only ~5%[27]. The performance of these phases relative to hydrogen embrittlement (HE), H trapping sites and IGCs in the Al alloys has been widely reported, and the elongation is greatly impaired in the H environment[28, 29]. Thereby, optimizing the corrosion resistance of Al alloys with given phases is difficult, and there continues to be a lack of research on the global optimization of corrosion resistance and mechanical elongation. Hence, it is urgently required to propose novel strategies to design new kinds of corrosion-resistant and high-elongation (anti-HE) Al alloys.

In this study, we illustrate the challenges for ML based on corrosion-mechanics data and propose new ML models combined with DFT calculations to design Al alloys having attractive corrosion resistance as well as mechanical properties. The accuracy of the new reinforcement self-learning model for corrosive elongation prediction archives 0.92 (goodness of fit, $R^2$). Besides, the formation energy, mechanical modulus, and the work function of the various potential phases are compared via DFT calculations. Subsequently, the optimized composition and preferred microstructure for attractive corrosion resistance and mechanical properties are recommended based on fused ML and DFT calculations approach. Finally, we successfully manufactured the Computation Designed Corrosion-Resistant Al (CDCR-Al) alloy and verified the reliability of the proposed design strategy. According to the slow strain rate testing (SSRT, in 0.1 M NaCl), the elongation of the CDCR-Al alloy is ~30%. In addition, its corrosion potential is also higher than -0.7 $V_{SCE}$. Multi-computations reveal that the design structures are favorable for trapping H atoms, and new GB exhibits higher cracking resistance than the raw η precipitated.

**Results**

*ML models for Al alloys with corrosion data*

More than 1,000 data points (including ~300 corrosive tensile data, Fig. S1a.) were applied to capture the effect of various elements on strength. For the strength prediction, the conventional Back Propagation Neural Network (BP-NN, Fig. 1a) models are still effective[30]. By adjusting the network depth, activation functions, evaluation functions, and learning rate, a conventional Al strength BP-NN model with $R^2$ as high as 0.967±0.007 is established within this study (Fig. 1b). All model accuracy is calculated based on the validation set (independent data, accounting for 20% of the total data). For the validation of the model generalization ability, the experimental UTS and SCC strength values of AA7005, commonly utilized in high-speed trains, are 311 MPa and 284 MPa[31], respectively. The predictions of the BP-NN model for AA7005 are 326 MPa (UTS, 4.99% error) and 287 MPa (SCC strength, 1.09% error), respectively (Fig. 1d).

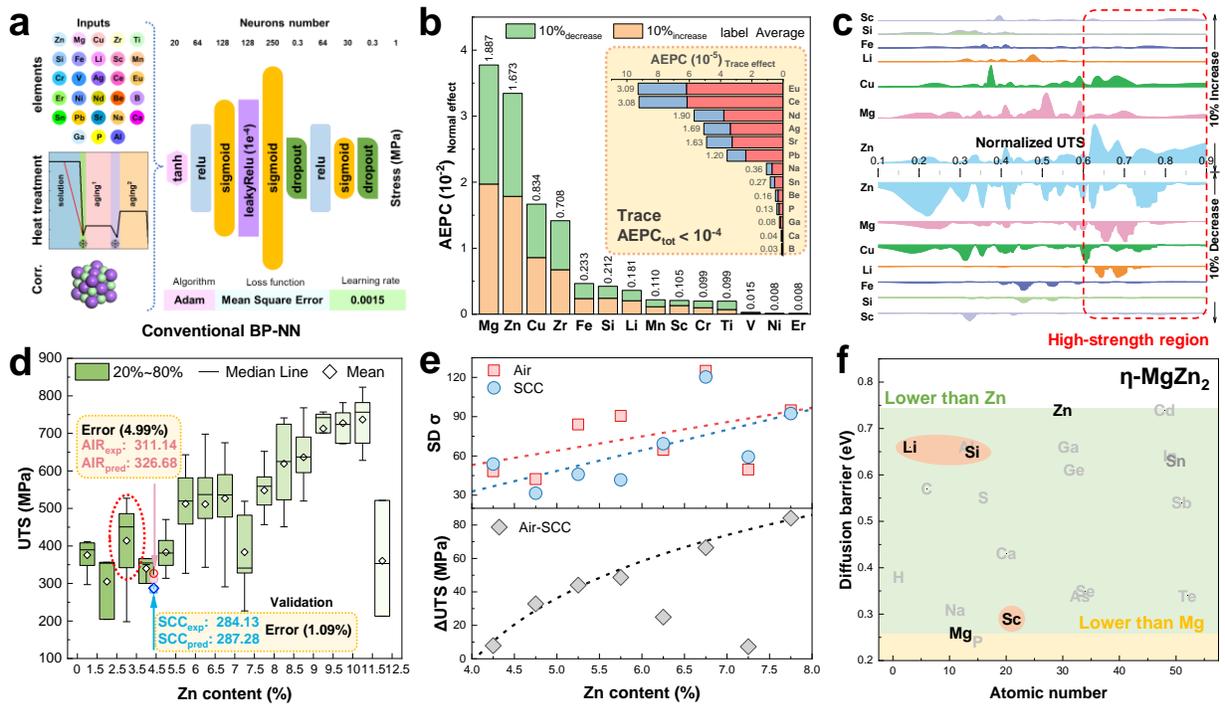

**Fig. 1. Conventional strength BP-NN model. a,** model structure; **b,** AEPC elements utility ranking. **c,** the strength contribution of elements whose AEPC is larger than 0.1. **d**, model verification and statistical relationship between Zn and UTS. **e,** SCC sensitivity of Al alloys with Zn content. Standard deviation $\sigma$ of data and strength loss $\Delta$UTS. **f,** the NEB-calculated diffusion barrier of elements which is lower than Zn.

Based on the strength BP-NN model, the area over element perturbation curves (AEPC, Eq. 1) are proposed to better distinguish the elemental contribution to Al alloy strength. This method integrates the conventional mean impact value and the area over the Most Relevant First perturbation curve (MoRF, used in visual identification)[32], however, it can greatly avoid the unfair influence brought by the amount of data. The hypothesis of applying this equation is the continuous relationship between before/after perturbed UTS (see the mathematical corollary in Fig. S2). The AEPC statistical results are shown in Fig. 1b. Despite excluding the effect of data amount, common alloying elements (Mg, Zn, and Cu) still have considerable impacts on strength. The improvement of Zr and Fe is more obvious than that of Si, while the average area of Sc and Er reaches $2.1\times10^{-3}$ and $1.6\times10^{-4}$ (area index), respectively. These overall area indices greater than $10^{-4}$ are considered to be prominent elements in Al alloys design.

$$AEPC = \frac{1}{L}\int_{y=0}^{L}\left|f\left(x_{\pm10\%}^{k}\right)-f\left(x_{raw}^{k}\right)\right|ydy \tag{1}$$

where *f(x)* indicates the predicted UTS via BP-NN, *x* represents the element content and its perturbation, *k* denotes the element type, *y* and *L* is the practical and the maximum strength in the Al alloy dataset, respectively.

It must be clarified that the AEPC masks the effects of elements in different strength intervals, not all attractive elements (area index $>10^{-4}$) are inclined to design high-strength Al alloy. Hence the unintegrated results are shown in Fig. 1c. The 2/3 (the normalized strength) is identified as the threshold value of high-strength Al alloys, it can be seen that the Zn has the most profound impact on the UTS. Admittedly, the synergistic alloying of Zn and Mg is the most effective method to produce the high-strength Al alloys (in Fig. S1a, the UTS of AA7xxx reaches ~800 MPa). Considering the excellent corrosion resistance of AA5xxx (Mg alloyed), the inferior corrosion resistance of AA7xxx is presumed to be related to Zn content.

The effect of Zn content on Al alloy strength is summarized in Fig. 1d, expectedly, the alloy strength significantly improves with Zn increases. However, the strength decreases noticeably when the Zn exceeds 12%. There are two anomalous regions which are lower than that of the surrounding region, namely the 2.5% to 3.5% range and the 7.0% to 7.5% range. In the latter case (7.0-7.5%), the mean UTS of Al alloys is only 384.3 MPa. Combined with their SCC properties (Fig. 1e), it is inferred that the increase of Zn content (<8%) aggravates the mechanical degradation of Al alloys. On the contrary, the AA7xxx with Zn content in the range of 3–4% has better mean UTS (~413.9 MPa), and there is also no great strength degradation reported. Therefore, the Zn range of 3–4% is a cost-effective corrosion-strength-designed Zn content range, which not only enhances the corrosion resistance of Al alloys but also reduces the density of Al alloys.

In addition, Cu, Li, Fe, and Sc also exhibit influence on the high-strength regions. The enrichment of Cu and Er at the GBs can improve the GB cohesive energy, which can reduce the occurrence of IGCs. Moreover, Zr in Fig. 1b likewise shows a significant effect on the strength of Al alloy, but it mainly refines grains[33] or promotes dispersed precipitation[34]. The fine grain strengthening increases the number of GBs which enlarges the risk of η/η' phases distributed at the GBs. Traditionally, Fe-containing phases are considered harmful components and are removed as much as possible during metallurgy[35], but ML shows that Fe has positive effects on the strength of Al alloy. Besides, Li is discarded owing to the DFT calculations described below.

According to elemental stability DFT calculations, Sc and Zn are overlapped in the diagram[13]. To avoid/modify the precipitation of η-$MgZn_2$ at the GBs, the diffusion barrier of elements which lower than Zn are shown in Fig. 1f. It can be found that diffusion ability of Sc is closed to Mg. Based on the above calculation, Sc (like Zn) and Er (improving the GB cohesive energy) are added to AA7005 to analyze their effects on mechanical and SCC properties (Fig. S1c). Compared with Er, the resulting improvement of Sc is more evident, particularly for UTS in the air. However, its excessive addition (>0.2 wt.%) does not hugely improve their SCC performance. Despite the trace Er cannot significantly improve the UTS, its addition leads to an incline towards anti-SCC.

Unlike the strength, the influence of the corrosive environment on the elongation is more prominent. Quantitatively, the highest elongation prediction accuracy of the conventional BP-NN algorithm is only 0.573±0.09. This impuissance is more obvious when predicting the elongation of high-strength Al alloys, this may be attributed to the fact that corrosion promotes premature cracking and merging of cracks resulting in elongation reduction. The high accuracy of hybrid strength-elongation ML models for strength masks its disadvantage in elongation predictions, especially in corrosive environments. Hence, we propose to use DFT calculations to accurately describe the elemental physical state in the Al, and develop a reinforcement learning algorithm with them (RL-NN, Fig. 2a) to separately predict the elongation. The calculated features include GB cohesive energy, matrix energy barrier, chemical potential (stability), adsorbed energy on (100) and (111), matrix substitution and interstice energy.

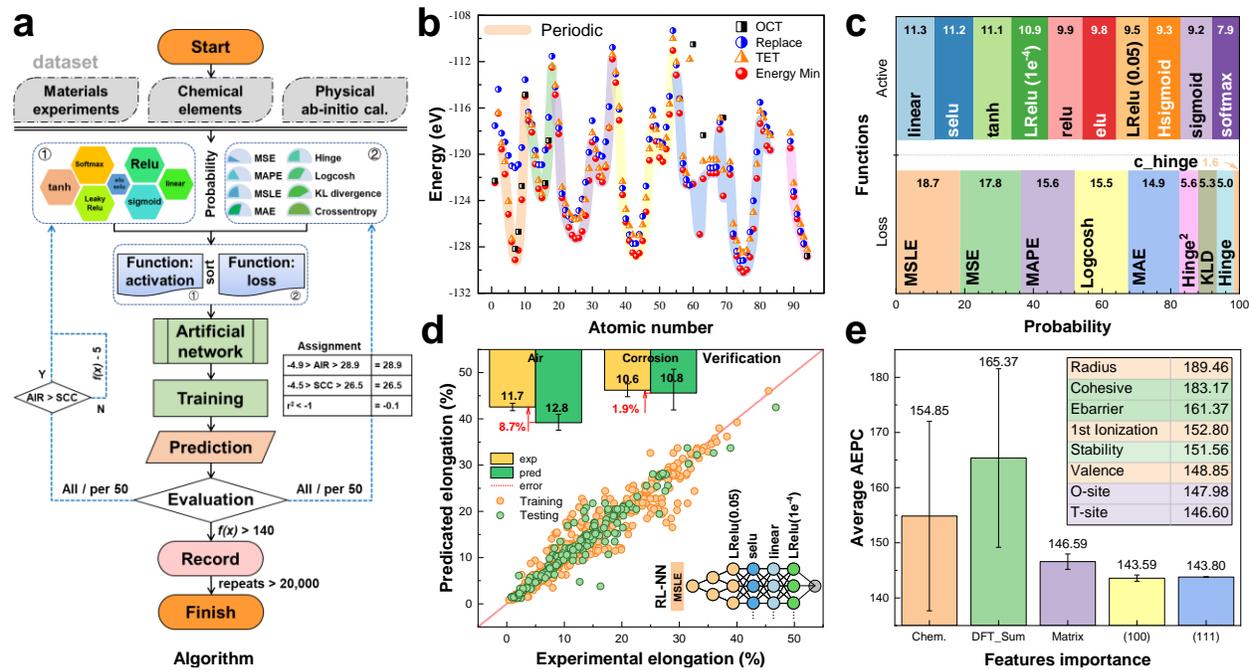

**Fig. 2. Reinforcement learning elongation model. a,** RL-NN algorithm. **b,** calculated matrix elemental energy. **c,** scoring of activation and loss function. **d,** model accuracy and verification. **e,** features importance comparison showing the contribution of calculations.

From the calculated matrix elemental energy (Fig. 2b), it can be easily found that the minimum energy exhibits periodic regularity. Clearly, when the atomic orbits of elements are half-filled ($p^3$ or $d^5$), their minimum energies are local lowest in each period. While the elements whose atomic orbits are fully occupied are the local highest (e.g., Er). In view of this law, the element absorbed energies on the Al (100) and (111) are statistically analyzed according to the atomic orbital occupation (Fig. S3a). It also clarifies that elements in the VA or VIIB group are easily adsorbed, while the adsorbed energy of noble gas elements with no empty atomic orbitals is greatly improved. When the outermost atomic orbital is not yet half-filled, the adsorbed energy of the atom decreases with the increase in the number of electrons in the outermost atomic orbital. The energy then increases after the atomic orbital is half filled. To predict the content of uncommon elements (sparse samples), all compositions are converted into chemical/physical innate characters which synthesize experimental and computational data (Fig. S4 and Table S2).

Except for the specialized dataset, the RL-NN model set eight activation functions and eight loss functions with the same initial accuracy probability. Then, several activation functions and one loss function are selected according to the probability, such that the optimal conventional neural network can be generated (the functions are randomly sorted). Each model is evaluated based on the equations in Fig. S3b, then the value is fed back to the accuracy probability. Statistically, the function probability fluctuation reveals which functions are more suitable for elongation predictions with corrosion tensile data.

As shown in Fig. 2c, the accuracy of the mean error functions (loss) is higher than the other types of functions. Unlike the commonly used mean square and mean absolute errors, mean squared log error (MSLE) is more suitable for the corrosion-mechanical Al alloy dataset. In terms of activation functions, the accuracy probability of the sigmoid function (commonly used in BP-NN) is exceptionally low at 9.18, ranking second to last and only higher than the softmax function. However, the activation functions, such as the linear, selu, tanh, and leakyRelu, have accuracy probabilities exceeding 10. Combining the best activation and loss functions, the model with an accuracy of 0.926±0.022 is developed (Fig. 2d). Its air elongation error is 8.7%, while the error in the corrosive environment is 1.7% (generalization ability validation). Combined with the BP/RL-NN, the integrated model has superior accuracy for the strength (0.96) and elongation (0.92) of Al alloys in air/corrosive environments.

Comparing the contribution of the traditional chemical and calculated features to the accuracy, Fig. 2e clearly indicates that the calculated features have higher positive influences on the RL-NN model. 5 of the top 8 features in the importance ranking are calculated features, which are GB cohesive energy, energy barrier, chemical potential, energy in T-site and O-site. Although the most related descriptor still is chemical atomic radius which the AEPC value reaches 189.4, that of GB cohesive energy is just 3.33% lower (183.1). Therefore, the elemental GB cohesive energies have profound impacts on the elongation prediction of Al alloys. Comprehensively, the average AEPC of the summarized calculated features (165) is greatly higher than that of the chemical features (155). After analyzing the calculated features, it also can be referred to that the calculated features related to the matrix have higher effects than the surface calculations. More specifically, the average AEPC importance of conventional calculated features is relatively low, such as that of adsorbed energy on the (100) and (111) is only 143.59 and 143.80, respectively.

*DFT calculation of the secondary phases assisting in microstructural screening*

While the added elements (ML determined) bring opportunities for the formation of new phases, they may also pose the risk of material degradation (mechanical properties and corrosion). Therefore, the formation energies and mechanical moduli of phases are analyzed within this study (Fig. 3a). This diagram shows calculated formation energies and modulus index (arithmetic square root of square sum of Young's modulus $E$ and shear modulus $G$, $\sqrt{E^2+G^2}$), and it is divided into 4 regions according to that of Al (short black dash line). The low value of formation energy indicates that phases can form in the matrix, while an extremely low formation energy value reflects the high possibility of phase growth which may weaken the intrinsic hardening effect[36]. To determine the shape and size of the phases more precisely, the interfacial energies of phases/matrix are required. Considering the excessive computational scale of interface calculations (factors including Miller indices, misorientations, coherent, and semi/non-coherent interfaces), the formation energies are prioritized. As for the mechanical hardening effects of phases, low modulus indices cannot be directly regarded as an absence of hardening effects, since the dispersion of phases may hinder the dislocation slip.

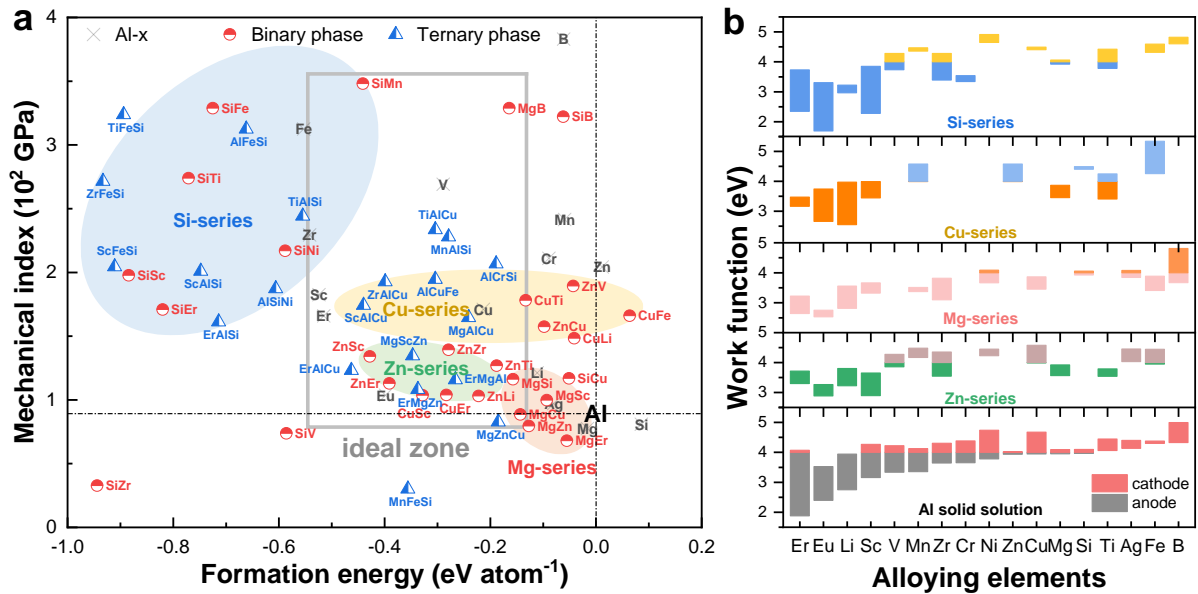

**Fig. 3. The phases DFT calculations related to mechanical properties and corrosion behavior. a,** mechanical moduli indices and formation energies. **b,** work functions. The elements Cu (MgZnCu, Imma), Er (ErMgZn$_2$, Fm$\bar{3}$m), and Sc (MgZn$_2$Sc, Fm$\bar{3}$m) can significantly improve the mechanical index of raw η-MgZn$_2$ precipitates.

Interestingly, the phases formed by the main alloying elements (Zn, Mg, Cu, and Si) appear to aggregate distinctly. The formation energy of the Mg group (pink) is in the range of -0.2 ~ 0 eV

atom$^{-1}$, but their mechanical indices are the weakest among the main alloying elements. The Zn group (green) forms more easily than the Mg group, and their moduli indices are also greater than Mg. Pure Si will not precipitate from the matrix (≈0.09 eV atom$^{-1}$), whereas its derivatives, having lower formation energy than Zn/Mg/Cu groups, such as β[37], B′[38], and Mg$_2$Si[39] can form and strengthen the alloys after heat treatment. Excessive coarsening of these Si-phases leads to a decrease in both mechanical properties[40] and corrosion resistance[41]. The Fe group exhibits the lowest formation energy among all alloying elements. Although the specific size of the phases is also governed by their interfacial energy, Fe is still unlikely to be the main alloying element due to experimentally detected harmful influence, despite its superior mechanical (Voigt statistical Young's modulus of AlFe is about 289 GPa)[42]. Therefore, the phases whose formation energies are lower than Fe (-0.55 eV atom$^{-1}$) may attenuate their mechanical strengthening effect and cause galvanic corrosion, therefore those phases should be avoided[43, 44].

Insight into Zr-induced grains refinement (GBs proliferate and favor the distribution of η/η' phases) and Fe/Si coarsening phases, their positions are set as the left boundary of the ideal zone (gray box in Fig. 3a). In addition, the accepted phases should have lower formation energy (right boundary of the gray box) and better strengthening effect (bottom boundary of the gray box) than the raw η/η' phases. To compensate for the mechanical loss (caused by the reduced Zn content), the upper left elements are more expected. Finally, the expected microstructure is to use ScAlCu, ErAlCu, or MgZnCu to replace or ameliorate the η/η' phases, simultaneously, Zr, V, and B are also added to make up for mechanical strength regress.

The corrosion behavior of the phases is another aspect to be optimized. Fig. 3b shows the work functions of various phases, and their positive and negative regions are identified by Al surface (111) which exhibits higher importance than (100) in the RL-NN model. Fundamentally, an excessively low work function tends to be subjected to pitting corrosion, while a high work function risks galvanic corrosion. From the work function in Fig. 3b Al solid solution, it can be seen that the selected Sc or Er is not conducive to the improvement of corrosion resistance, whose lowest work function are 3.16 eV and 1.88 eV respectively. This warns that the growth of Sc/Er-related phases must be inhibited. Besides, Fe can improve the work function of Al matrix (4.30 eV). However, given the lowest formation energy of Fe group, superfluous Fe will form large-sized cathode particle resulting in galvanic corrosion[45]. As for Li, its combination with any other element generates phases with extremely low work function (2.75 eV), which are also fatal to corrosion resistance.

The work function of the Zn or Mg groups is narrow, regardless of whether the highest/lowest work function is closer to that of the Al matrix. Whereas, the phases formed by Cu or Si are difficult to be solely classified as cathode or anode, where the work function range is relatively large[44]. Meanwhile, Sc and Er approved by the aforementioned BP/RL-NN model can improve their lowest work functions by doping with Cu, which is 9.15% and 68.37% respectively[46]. Therefore, it can be inferred from calculations that Zn and Mg are less harmful, Fe, Sc, and Er must be precisely added to avoid the formation of the harmful phases, and Li is completely discarded. Partial Cu is expected to combine with Sc/Er to prohibit the pitting corrosion and the other Cu is desired to diffuse into the GBs.

*Design and performance evaluation of Al alloy*

Based on the identified element, the trained BP/RL-NN models and genetic algorithm are used to determine element compositions. The optimization goal is synergetic UTS and elongation in corrosive environments. Considering the high dimensionality and extremely wide composition range for the genetic algorithm (full-space search), we perform optimization for AA7005 (Table

S3). The strategy is to reduce Zn and increase Cu, Fe, and Ti. Simultaneously, small amounts of Sc/Er (corrosion-related) and trace amounts of Zr, V, and B (mechanical property-related) are added. The composition restriction in Fig. 4a is divided into four gradients, namely main alloying (broadly, Zn, Mg, Cu, and Si), important but need to be strictly limited (<0.5%, Fe), trace (<0.3%, Sc, Mn, Ti, and Er) and sub trace (<0.1%, Zr, V, and B).

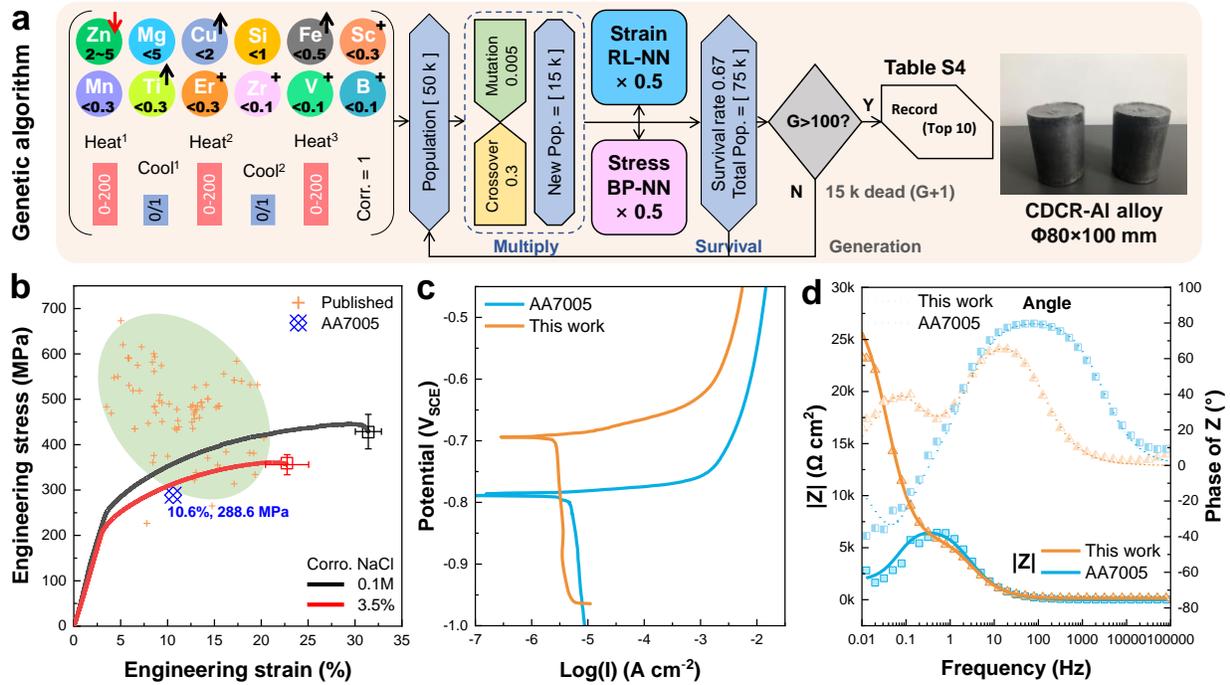

**Fig. 4. Properties evaluation of the CDCR-Al alloy. a,** composition limitation and heat treatment determined by genetic algorithm. **b,** slow strain rate testing of CDCR-Al alloy, raw AA7005 and reported works in corrosive environments. Compared with the AA7005, the increase percent of stress corrosion strength and failure elongation of CDCR-Al alloy is 53% and 193%, respectively. **c,** potentiodynamic polarization curve. **d,** Bode electrochemical impedance spectroscopy.

After 100 generations, the composition of the optimal alloy (Table S4) converges, even in a high probability mutation environment (3%). The initial unlimited heat treatment determined is 0.80, 1, 0.12, 1, and 0.41, respectively. Given the diffusion barrier (0.57 eV)[47] and activation energy (3.2 kcal mol$^{-1}$)[48] of Cu in Al, the implemented solution treatment is ~505°C for 8 h to promote the Cu diffusing into the GBs or combining with Sc. This temperature reconciles the conventional solution temperature of Al-Zn-Mg alloys (~475°C) and Al-Cu alloys (~530°C)[49, 50]. Besides, the aging process is consistent with that of Al-Zn-Mg alloys, with a temperature of 120°C for 4 h. Short aging time is also beneficial to avoid excessive formation of the η/η' phases and more stable Al-Sc phase (Cu free). Subsequently, the CDCR-Al alloy is successfully manufactured via electromagnetic stirring and die casting, producing a billet with a size ϕ80×100 mm (Fig. 4a).

Employing a high solid solution temperature (505°C for 8 h), the diffusion of Cu with relatively high diffusion barrier is more active, such that the partial $Al_3Sc$ phases are doped with Cu. To verify the CDCR-Al alloy resistance to SCC, SSRTs ($10^{-6}\,s^{-1}$) were carried out in corrosive environments (0.1 M NaCl and 3.5 wt.% NaCl), with results shown in Fig. 4b. Clearly, the CDCR-Al alloys in a moderately corrosive environment displayed fine SCC resistance, with UTS reaching 443.3±38 MPa and elongation reaching 31.4±1.4%. This exceeds the reported results (Fig. 4b, green region). In terms of the most pronounced elongation reduction in SCC, the elongation of the CDCR-Al alloy is almost double that of the AA7005 currently in service, and the verification tests are summarized in the Fig. S5.

The potentiodynamic polarization curve and electrochemical impedance spectroscopy (EIS) of the CDCR-Al alloy, AA7005, AA6005, and AA5083 are tested to evaluate their corrosion behavior. As shown in Fig. 4c, the corrosion potential ($E_{corr}$) of the CDCR-Al alloy is significantly improved from the AA7005 (+0.095 $V_{SCE}$, Saturated calomel electrode), its value (-0.694 $V_{SCE}$) is, in fact, closer to AA6005. The CDCR-Al alloy is similar to AA5083 at high potential, which is also smaller than AA7005. Unfortunately, not only exhibiting the lowest $E_{corr}$, AA7005 retains the largest $i_{corr}$. Once the AA7005 is in the anodic region, its $i_{corr}$ surges to 1.24 mA $cm^{-2}$. Combined with Fig. S6b, it can be seen that AA7005 had an inductive reactance phenomenon at low frequency. This is attributed to the difficulty of diffusion of metal ions or other conductors. The CDCR-Al alloy exhibits two time-constants in the intermediate and low frequency. Its film impedance at the low frequency reaches 23.12 kΩ, illustrating that the surface oxide film has excellent protection (the capacitance index ~0.916). Hence, under the same environment, the protection provided by the surface film on the CDCR-Al alloy outperforms that of AA7005.

*Microstructure observation and performance explanation*

From Fig. 5a, it is observed that there are numerous and fine phases dispersed in the grains which an average size is 36.62 nm. Combined with the energy dispersive spectroscopy (EDS) result, Al, Sc, and Cu are found. According to the formation energy shown in Fig. 3a, the $Al_3Sc$ exhibits lower formation energy than the Al-Sc-Cu phases, which indicates the $Al_3Sc$ phases are more stable. To confirm whether the Cu doping in the $Al_3Sc$ phases, a three-dimensional atom probe (3DAP) is performed and further confirmed that a small amount of Cu was doped into the $Al_3Sc$ interior (Fig. 5c). It is found from Fig. 5b that the Al-Sc-Cu phase shows coherent relationship with the Al matrix, with $[200]_{Al}//[100]_{prep}$ and $[0\bar{1}1]_{Al}//[0\bar{1}1]_{prep}$. In addition, the interplanar spacings of phases are accurately measured as 3.86 Å and 2.77 Å, respectively. However, the length of the primitive $Al_3Sc$ cell is confirmed as 4.105 Å[51]. When Cu replaces one Al atom in the $Al_3Sc$ cell, the cell length can be reduced to 3.86 Å. Therefore, combined with the EDS, high-resolution transmission electron microscopy (HRTEM), and DFT calculations, we reckoned that particle $Al_3Sc$ are doped with Cu atoms.

Furthermore, the Gibbs free energy change Δ*G* of H in the Al matrix and phases are computed and summarized in Fig. 5d. It can be clearly seen that the H captured ability of raw η/η′ phases is weakest (0.05 eV $H^{-1}$), which is greatly higher than that in the Al matrix. Thereby, the inner distributed η/η′ phases cannot limit the diffusion of H in grains. When the $Al_3Sc$ phases are formed in the grains, the Δ*G* decreases to 0.01 eV $H^{-1}$ which is still slightly higher than that in the matrix.

However, once Cu atoms doping into the Al$_3$Sc, the interstice site of Al-Sc-Cu phases is greatly reduced to -1.44 eV H$^{-1}$. Therefore, this modification reduces the lattice parameters of phases and simultaneously enhances their H-trapping ability.

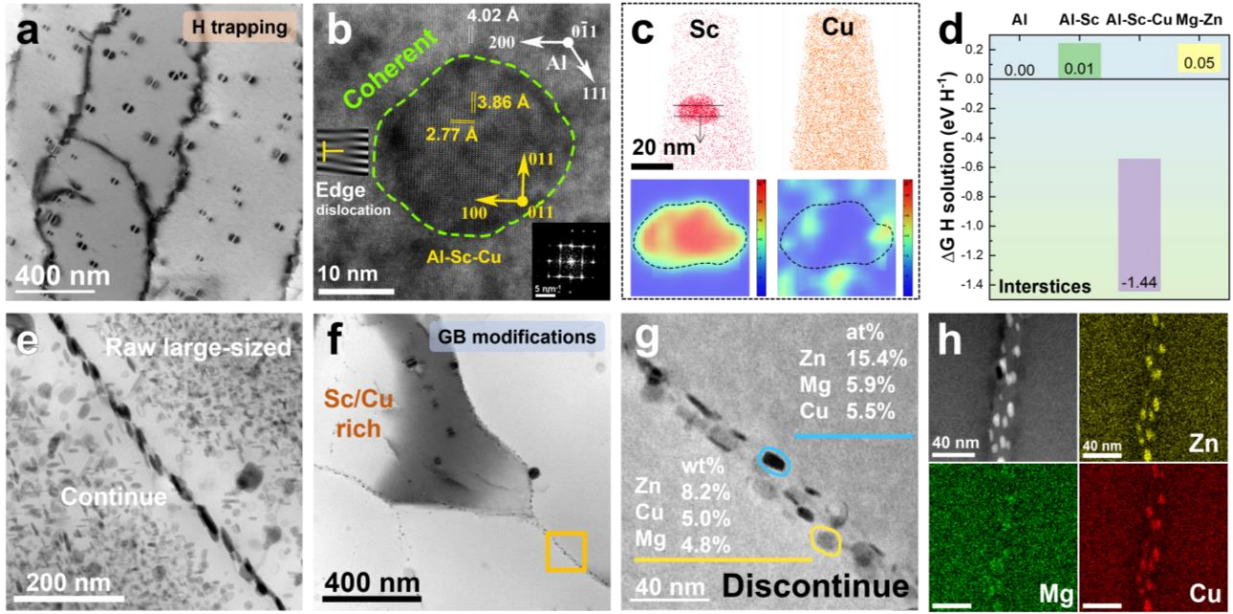

**Fig. 5. Microstructure of the CDCR-Al alloys.** (**a-d**) H trapping structures inside the grains and (**e-h**) GB precipitation transformation. **a,** dispersed fine precipitates inside the grains. **b,** interfacial orientation relationship and Fast Fourier Transform for precipitates. **c,** 3DAP maps of precipitates with Cu doping. **d,** Gibbs free energy change for H trapping. **e,** raw continuous distributed large-sized η phase at the GB. **f-g,** discontinuous small-sized GB precipitations. **h,** EDS mappings.

The segmented and modified GB η/η' phases (Fig. 5e-g) also significantly improve the anti-SCC performance of the CDCR-Al alloy. As shown in Fig. 5e, the raw continuously distributed large η/η' phase at the GBs is significantly sensitive to the cracking, whose GB cohesive energy is only 13.4% of raw Al GBs. However, employing the ML method, the size of the precipitation inside the GBs is decreased to 10.87±2.67 nm (Fig. 5g). Furthermore, the η/η' phases in the GBs are doped with Cu which are transformed into the Zn-Mg-Cu phases shown in Fig. 5h, Cu can enhance the GB cohesive energy (0.258 eV for Σ7(111) GB). Based on the GB HRTEM results, it is found that the precipitates continue to show a coherent relationship with the Al matrix.

It must be reiterated, however, that not all GBs are completely transformed into new structures, such as the partial GBs in Fig. 5a. This may be a reason for the 8.6% decrease in elongation of the CDCR-Al alloy in harsh corrosive environments. Undesirably, Er does not apparently segregate into the GBs but dissolves into the Al matrix. Except for these, trace Er is found to combine with Sc (Fig. S10). Pitting is frequently caused by the phases with low work functions, we therefore utilize a scanning Kelvin probe force microscope (SKPFM) to observe the Volta potential difference of the phases (Cu and Sc rich) in the CDCR-Al alloy (Fig. S9). Fortunately, the Volta difference in this case is 63.25 mV, meaning the possibility of galvanic corrosion is low.

**Discussion**

Under the same corrosive environment, the corrosion current of CDCR-Al is close to that of AA5xxx, with a surface oxide film that is intact without micro-pitting. The average work function of potential ternary phases are shown in Fig. 6a, showing the raw Al-Sc phase exhibits extremely low work function (3.63 eV). However, when the Cu is doped to form the Al-Sc-Cu phases, the work function is greatly improved (4.18 eV), which is slightly larger than that of the Al matrix. Moreover, we utilize the SKPFM observing the height (Fig. 6b) and Volta potential difference (Fig. 6c) of nano-scale Al-Sc-Cu phase. It can be confirmed that the Volta potential difference of Cu doped $Al_3Sc$ is extremely close to the Al matrix, which is only 14.6 mV.

As for the complete protective film, this is attributed to the addition of Sc and Er. These two elements have stronger binding abilities to O than Al, their calculated formation energies are -3.58 eV ($Sc_2O_3$) and -3.71 eV ($Er_2O_3$), respectively. Besides, band structure calculations show that their band gaps ($E_g$) decrease with the enhancement of formation ability (Fig. S11). All $E_g$ of the oxides are greater than 4.5 eV, resulting in an insulator property.

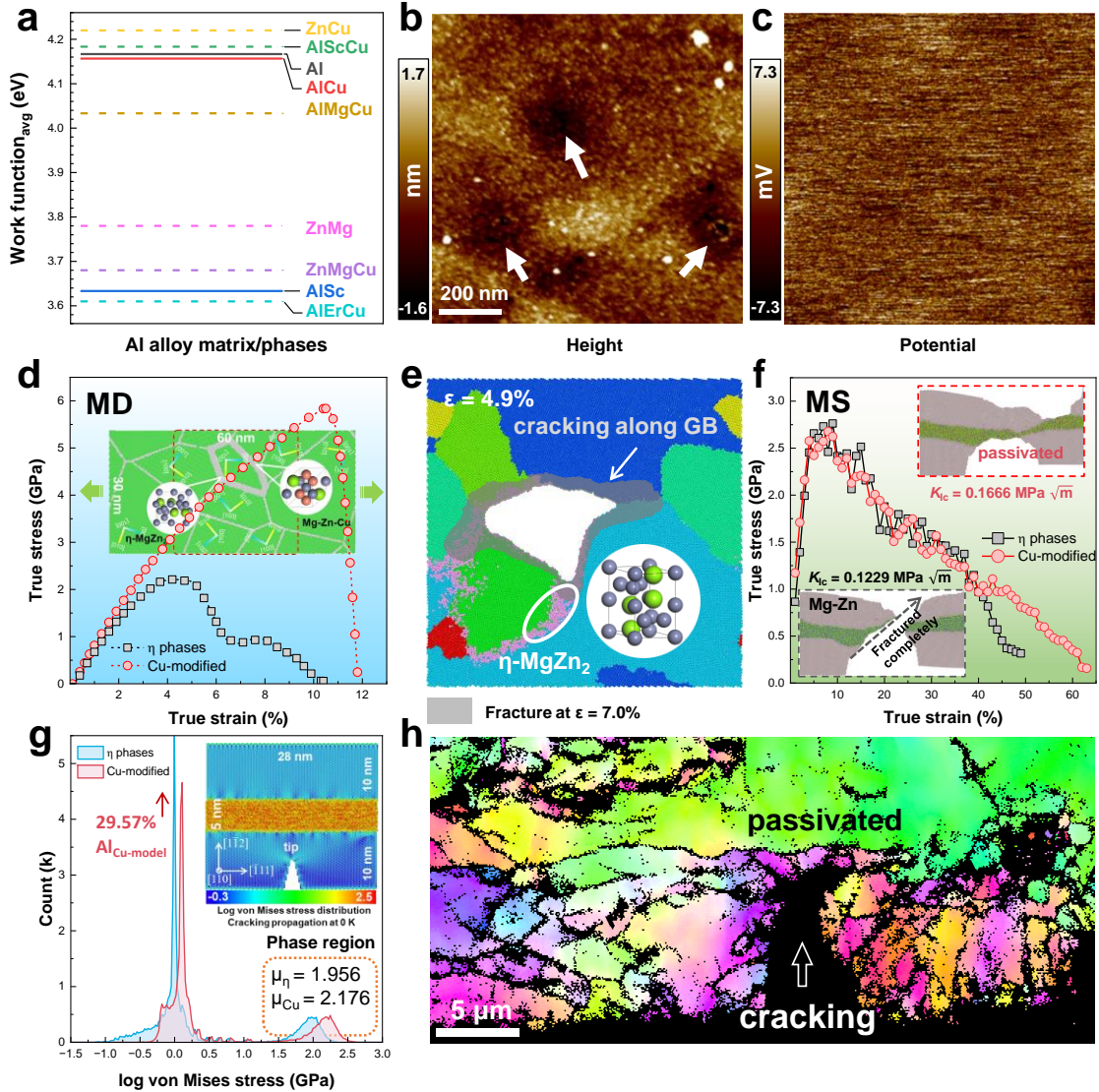

**Fig. 6. Mechanism interpretation of the CDCR-Al alloy. a,** average work functions of phases. The difference of Al-Sc-Cu phases in **b,** height and **c**, potential detected by SKPFM. **d,** stress-strain curve of polycrystalline MD tensile simulations. **e,** Failure morphologies of η models with strain performed. The gray region is the failure with a strain of 7%. **f,** stress-strain curve of MS cracking simulations. **g,** von Mises stress comparison of cracking models. μ is mathematical expectation of the Gaussian distribution. **h,** EBSD morphology of SSRT secondary crack.

To elaborate the mechanism of Cu modified η phases to cracking, a dataset (containing 10,122 DFT configurations) is used to train an Al-Mg-Zn-Cu interatomic potential. As shown in Fig. S12, its energy and force accuracies reach 6.50 meV atom$^{-1}$ and 22.19 meV Å$^{-1}$ respectively, which are lower than the report work[52]. Two kinds of models are used to distinguish the cracking modes of Al alloys. The polycrystal models (Fig. 6e) are modified with precipitates at an annular GB, one represents the η phases and the other is the Mg-Zn-Cu phase. From MD stress-strain curves at 300 K, it can be found that the strength of Cu-modified polycrystalline model (~5.83 GPa) is higher than η phases (~2.21 GPa). Besides, the cracking mode of η phases model is also different from that of the Cu-modified GB model. When the strain reach 4.9%, there is a hole near the GBs, with the strain performing, it apparently propagates along the GBs at a strain of 8.0% (gray area in Fig. 6f), while the Cu-modified model remains intact whose elongation reaches 11.8% (Fig. S12).

When one GB with a crack tip (molecular static, MS) is considered, the strength of Cu-modified model at 0 K is like the η phases model, but its failure elongation (63%) is still better than that of η phases (49%). Computing the von Mises stress (Fig. 6e), we can find that not only the Cu-modified phases (MgZnCu phase in Fig. 5h) in GB, but the stress of Al matrix (near GB) is also approximately 29.57% greater than that of η phases models. Different from the normal cracking simulation, not the crack tip, the high stress at the GB trigger the dislocation emission. Whereas the initial dislocation density of Cu model is higher than the η phases models, although it owns less stacking fault atoms. During the cracking, the increase rate of dislocation density of Cu model is greatly larger than that of η phases models, it indicates that the Cu-modification hinders the slippage of dislocations. This difference also affects the critical fracture toughness $K_{Ic}$[53], that of Cu-modified model (0.1666 MPa m$^{-1/2}$) is ~35.56% greater than the η phases model, indicating the crack resistance of Cu-modified GB is stronger. Moreover, the crack in η phases model propagates more easily, it is fractured completely at a strain of 49% in Fig. 6g. While the tip of Cu-modified GB is apparently passivated.

To detect the cracking mode and verify the effect of Zn-Mg-Cu phases on the GBs, the propagation of secondary cracks in the CDCR-Al alloy after SSRT is observed by EBSD (Fig. 6h). Except for the primary crack, it is difficult to find secondary cracks (<20 µm) in the alloy. Certainly, the identification rate of the sample edge is low due to the combined effect of deformation and corrosion. Combined with the SEM-detected crack position, it is found that the secondary crack in the CDCR-Al alloy propagates in a transgranular manner.

In conclusion, we propose a novel strategy combining ML and high throughput ab-initio calculations to efficiently design an Al alloy with excellent corrosion resistance and mechanical properties. The designed structure Al-Sc-Cu nanoprecipitates (in grains) and Cu modified η/η' phases (at the GBs) have been proven to be beneficial to the corrosion resistance improvement of Al alloys. The CDCR-Al alloy exhibited excellent mechanical properties (UTS ~450 MPa and its elongation exceeded 30% in the 0.1 M NaCl) and corrosion potential (-0.694 $V_{SCE}$). The computational results demonstrate the Cu-doped Al-Sc phases can greatly capture the H atoms, simultaneously, the Cu-modified GB exhibits stronger resistance to cracking. Despite not all GB segregations are transformed into designed structures. It is worth affirming that the use of high-throughput ab-initio calculations has greatly avoided the formation of harmful phases, thereby improving the efficiency of alloy design.

**Methods**

The basic ML algorithms utilized are BP-NN and deep learning API Keras. The original Al mechanical data comes from the published papers, these testing comply with the standard GB/T 228, ISO 6892 or ASTM E8 to ensure data quality. During data cleaning, data similarity and element types are checked to eliminate the same data and small sample elements (<5, e.g., Y, Hf, Nb, Cd, and La). The dataset size exceeds 1000, and the training and testing data ratios are 0.8 and 0.2, respectively. Fig. 1a describes the optimizer (Adam), loss function (MSE), learning rate (0.0015), activation functions, and the neurons number of strength BP-NN model. The total epoch of strength model training is 20,000, and layers 6 and 9 are set with a dropout parameter of 0.3 to prevent the BP-NN from overfitting. The features element and mechanical properties are linearly normalized. For the BP-NN model, the final dataset contains 34 input features, namely Zn, Mg, Cu, Zr, Ti, Si, Fe, Li, Sc, Mn, Cr, V, Ag, Ce, Eu, Er, Ni, Nd, Be, B, Sn, Pb, Sr, Na, Ca, Ga, P, Al, heat1, cool1, heat2, cool2, heat3, and corrosion. The entire heat treatment is divided into 5 components (3 for heating and 2 for cooling). For the cooling process in heat treatment (the corrosion feature is similar), the feature is set to 1 if the event exists, and 0 otherwise. Learning from the Arrhenius equation, the heating process is transformed into atomic diffusion (Eq. 2).

$$heat = t\exp(T+273)^{-1} \qquad (2)$$

where $T$ is the Kelvin temperature (K), and $t$ indicates the holding time (10 h).

The genetic algorithm initially generates random 50,000 alloys within the composition limitation. These alloys are multiplied and mutated to form 15,000 new alloys. Each alloy (75,000 in total) is given a survival probability by the BP/RL-NN models (predicated strength and elongation), then 15,000 alloys are eliminated according to their probabilities.

Due to the inferior corrosive elongation prediction of the conventional model, the elements combined with their contents are converted into chemical features, i.e., mass, atomic radius, electronegativity, 1$^{st}$ ionization energy, and valence electrons. Besides, the physical features calculated by the ab-initio method, such as GB cohesive energy, diffusion, dissolution energy in the Al matrix, and adsorption energy on the Al$_{(100)}$ and Al$_{(111)}$, are utilized to enrich the dataset. The data features and their correlation analysis are shown in Fig. S4.

All ab-initio calculations (9,704 in total) are performed using Vienna Ab-initio Simulation

Package (VASP 5.4)[54]. The generalized gradient approximation is applied with the Perdew-Burke-Ernzerhof (PBE) exchange-correlation functional[55, 56]. The dissolution and surface adsorption Al models are summarized in Fig. S13. Specifically, the dissolution sites include substitutional, octahedral, and tetrahedral interstitial sites. The adsorption models contain bridge, top, and HCP sites on the Al$_{(100)}$, and bridge, top, HCP and FCC on Al$_{(111)}$. The electron and force convergence accuracy are set to $10^{-5}$ eV and 0.01 eV Å$^{-1}$, respectively. The setting of *k*-points is based on the spatial length (0.2 Å$^{-1}$) in each direction, and the cutoff energy is 450 eV. For mechanical modulus calculations, 79 kinds of phases in Al alloys are considered (Fig. S14). The Voigt models are used to synthesize the elastic constants and generate Young's, bulk, shear modulus, and Poisson's ratio[57]. The detailed calculations are based on Eqs. 3 and 4.

$$E = WV \tag{3}$$

$$c_{ij} = \frac{\partial W}{\partial \varepsilon_i \partial \varepsilon_j} = \frac{1}{V} \frac{\partial^2 E(V, \varepsilon_k)}{\partial \varepsilon_i \partial \varepsilon_j} \tag{4}$$

where *E* is the ab-initio calculation total energy, *W* is the cohesive energy density, and *V* expresses the model volume. $c_{ij}$ and $\varepsilon_{ij}$ are the elastic constant and strain components, respectively.

The work function differences ($\psi_m^p$) represent the corrosion tendency of phases in Al alloys, it can be calculated by Eqs. 5 and 6.

$$\psi = \phi - E_f \tag{5}$$

$$\psi_m^p = \psi_p - \psi_m \tag{6}$$

where $\phi$ indicates the vacuum level determined from the average potential in the vacuum gap. $E_f$ is the Fermi level. $\psi_p$ ($\psi_m$) is the work function of the phase (Al matrix).

The band structures of Al$_2$O$_3$, Sc$_2$O$_3$, and Er$_2$O$_3$ are obtained using Heyd-Scuseria-Ernzerhof (HSE06)[58], which utilizes non-local exact Hartree-Fock (HF) exchange mixed with PBE to describe exchange-correlation energy ($E_{XC}^{HSE}$, Eq. 7).

$$E_{XC}^{HSE} = E_X^{HF,SR}(\omega) + (1-\alpha)E_X^{PBE,SR}(\omega) + E_X^{PBE,LR}(\omega) + E_C^{PBE} \tag{7}$$

where the short-range (*SR*) component includes the HF energy $E_X^{HF,SR}$ and PBE energy $E_X^{PBE,SR}$, while the long-range (*LR*) mainly is the PBE term $E_C^{PBE}$. $\alpha$ indicates the percentage of HF energy and $\omega$ is the range separation between SR and LR.

Cracking MD/MS simulations are using Large-scale Atomic/Molecular Massively Parallel Simulator[59]. The MD system contains ~440,000 atoms, while the MS system is about 35,600 atoms. All models are three dimensions. The basic ML interatomic potential method (Moment tensor potential, MTP) is developed by Shapeev groups[60], and the level of MTP is 16 with energy-weight (1), force-weight (0.01), and stress-weight (0.001). All configurations are randomly divided into training set (0.8) and validating set (0.2). Given sufficient raw configurations, active learning is performed only 2 times (deforming at 800 K). The isothermal-isobaric ensembles (MD models) are initialized with the temperature and pressure by velocity scaling method to 300 K and 0 MPa respectively. During deformation, the systems are maintained at 300 K, and a strain of 0.1% is applied to the systems along the *z*-axis every 1,000 steps. As for MS models, the crack boundary is non-periodic and shrink-wrapped. Every step the *y*-direction (perpendicular to the cracking direction) is stretched 1%, and then the cg and fire methods are used to minimize. The critical fracture toughness $K_{Ic}$ and von Mises stress $\sigma_v$ are calculated according to Eqs. 8 and 9.

$$K_{Ic} = 1.122\sigma\sqrt{\pi a} \tag{8}$$

where $\sigma$ is the uniaxial stress and $a$ is the crack length.

$$\sigma_v = \sqrt{\frac{(\sigma_{xx}-\sigma_{yy})^2 + (\sigma_{yy}-\sigma_{zz})^2 + (\sigma_{zz}-\sigma_{xx})^2 + 6(\sigma_{xy}^2 + \sigma_{yz}^2 + \sigma_{xz}^2)}{2V^2}} \tag{9}$$

where $\sigma_{ij}$ is the stress tensor and $V$ indicates the atomic volume.

Scanning electron microscopy (SEM, S-3400N, Hitachi, Japan) combined with electron backscattered diffraction (EBSD) is used to detect the fracture morphology and grains information (size and orientations). The EBSD samples are electron-polished in a mixed solution of 20 vol.% perchloric acid and 80 vol.% ethanol with a voltage of 18 V for 30 s. An FEI Talos F200X TEM is performed to observe the structure of precipitates and energy-dispersive X-ray spectroscopy. The 3DAP samples are fabricated by the focused ion beam (FEI Helios Nanolab 600i)/SEM, then they are detected in the laser mode with a pulse repetition rate of 200 kHz, and specimen temperature is 50 K. The dimensions of the mechanical tensile samples with 3 mm thickness are shown in Fig. S5b. For corrosion resistance estimations, a three-electrode system (working electrode, saturated calomel electrode, and Pt counter electrode) is used to carry out the open-circuit potential, EIS, and potentiodynamic polarization measurements. The size of the working electrodes is 10×10×3 mm, and all samples are polished with 0.25 µm pastes. The EIS frequency varies from 100 kHz to 10 mHz, and the scanning rate is 10 mV min$^{-1}$ during potentiodynamic polarization measurements. The surface Volta potential difference is detected by the SKPFM (Multimode 8, Bruker). The scanning rate of SKPFM is 0.5 Hz with a resolution of 512×512.

**Acknowledgments**

This research was supported by the National Natural Science Foundation of China (No.52125102), Fundamental Research Funds for the Central Universities (No. FRF-TP-20-01B2), Guangdong Basic and Applied Basic Research Foundation (No. 2020B1515120093) and Special Fund Support for Taishan Industrial Leading Talents Project. Yucheng Ji thanks for the support from China Scholarship Council (#202106460037). The authors acknowledge the platform support of the National Supercomputer Center in Tianjin and USTB MatCom of Beijing Advanced

Innovation Center for Materials Genome Engineering.

**Author Contributions**

Y.J. compiled high-throughput calculation and machine learning algorithm, performed the experiments verification and wrote the manuscript. Y.J. and C.D. conceptualized the project and designed the research. X.F. collected and pre-processed the training data. F.D., Y.X., Y.H., M.A., and F.X. analyzed the experiment data and edited the manuscript. D.C. simulated the band structures of oxides. P.D. analyzed the ab-initio calculation result and edited the manuscript. J.R. optimized the machine learning algorithm. X.L. helped fine-tune the project.

**Data Availability Statement**

The data that support the findings of this study are available from the corresponding author upon reasonable request. The critical machine learning code and dataset are publicly available on https://github.com/yucheng-ji/CDCR-Al.

**Conflict of Interest**

The authors declare no conflict of interest.

**References**

1. Zhang X, Chen Y, Hu J. Recent advances in the development of aerospace materials. *Prog Aerosp Sci* 2018, **97:** 22-34.
2. Sun X, Han X, Dong C, Li X. Applications of aluminum alloys in rail transportation. *Adv Al Compos Alloys* 2021, **9:** 251-268.
3. Benedyk JC. Aluminum alloys for lightweight automotive structures. In: Mallick PK (ed). *Materials, Design and Manufacturing for Lightweight Vehicles*. Woodhead Publishing, 2010, pp 79-113.

4. Li X, Hansen V, Gjønnes J, Wallenberg L. HREM study and structure modeling of the η′ phase, the hardening precipitates in commercial Al-Zn-Mg alloys. *Acta Mater* 1999, **47**(9)**:** 2651-2659.

5. Sun W, Zhu Y, Marceau R, Wang L, Zhang Q, Gao X*, et al.* Precipitation strengthening of aluminum alloys by room-temperature cyclic plasticity. *Science* 2019, **363**(6430)**:** 972-975.

6. Garner A, Euesden R, Yao Y, Aboura Y, Zhao H, Donoghue J*, et al.* Multiscale analysis of grain boundary microstructure in high strength 7xxx Al alloys. *Acta Mater* 2021, **202:** 190-210.

7. López Freixes M, Zhou X, Zhao H, Godin H, Peguet L, Warner T*, et al.* Revisiting stress-corrosion cracking and hydrogen embrittlement in 7xxx-Al alloys at the near-atomic-scale. *Nat Commun* 2022, **13**(1)**:** 1-9.

8. Ji Y, Dong C, Wei X, Wang S, Chen Z, Li X. Discontinuous model combined with an atomic mechanism simulates the precipitated η′ phase effect in intergranular cracking of 7-series aluminum alloys. *Comput Mater Sci* 2019, **166:** 282-292.

9. Lalpoor M, Eskin D, ten Brink G, Katgerman L. Microstructural features of intergranular brittle fracture and cold cracking in high strength aluminum alloys. *Mater Sci Eng, A* 2010, **527**(7-8)**:** 1828-1834.

10. Liu LL, Pan QL, Wang XD, Xiong SW. The effects of aging treatments on mechanical property and corrosion behavior of spray formed 7055 aluminium alloy. *J Alloys Compd* 2018, **735:** 261-276.

11. Wang Y, Sharma B, Xu Y, Shimizu K, Fujihara H, Hirayama K*, et al.* Switching nanoprecipitates to resist hydrogen embrittlement in high-strength aluminum alloys. *Nat Commun* 2022, **13**(1)**:** 6860.

12. Stemper L, Tunes MA, Tosone R, Uggowitzer PJ, Pogatscher S. On the potential of aluminum crossover alloys. *Prog Mater Sci* 2022, **124:** 100873.

13. Ji Y, Dong C, Chen L, Xiao K, Li X. High-throughput computing for screening the potential alloying elements of a 7xxx aluminum alloy for increasing the alloy resistance to stress corrosion cracking. *Corros Sci* 2021, **183:** 109304.

14. Gupta RK, Deschamps A, Cavanaugh MK, Lynch SP, Birbilis N. Relating the early evolution of microstructure with the electrochemical response and mechanical performance of a Cu-rich and Cu-lean 7xxx aluminum alloy. *J Electrochem Soc* 2012, **159**(11)**:** C492.

15. Fang H, Chao H, Chen K. Effect of Zr, Er and Cr additions on microstructures and properties of Al–Zn–Mg–Cu alloys. *Mater Sci Eng, A* 2014, **610:** 10-16.

16. Hart GL, Mueller T, Toher C, Curtarolo S. Machine learning for alloys. *Nat Rev Mater* 2021, **6**(8)**:** 730-755.

17. Zhou Z, Chen K, Li X, Zhang S, Wu Y, Zhou Y*, et al.* Sign-to-speech translation using machine-learning-assisted stretchable sensor arrays. *Nat Electron* 2020, **3**(9)**:** 571-578.

18. Coelho LB, Zhang D, Van Ingelgem Y, Steckelmacher D, Nowé A, Terryn H. Reviewing machine learning of corrosion prediction in a data-oriented perspective. *npj Mater Degrad* 2022, **6**(1)**:** 8.

19. Nash W, Zheng L, Birbilis N. Deep learning corrosion detection with confidence. *npj Mater Degrad* 2022, **6**(1)**:** 26.

20. Xue D, Balachandran PV, Hogden J, Theiler J, Xue D, Lookman T. Accelerated search for materials with targeted properties by adaptive design. *Nat Commun* 2016, **7**(1)**:** 1-9.

21. Zhao Q, Yang H, Liu J, Zhou H, Wang H, Yang W. Machine learning-assisted discovery of strong and conductive Cu alloys: Data mining from discarded experiments and physical features. *Mater Des* 2021, **197:** 109248.

22. Wen C, Zhang Y, Wang C, Xue D, Bai Y, Antonov S*, et al.* Machine learning assisted design of high entropy alloys with desired property. *Acta Mater* 2019, **170:** 109-117.

23. Feng X, Wang Z, Jiang L, Zhao F, Zhang Z. Simultaneous enhancement in mechanical and corrosion properties of Al-Mg-Si alloys using machine learning. *J Mater Sci Technol* 2023, **167:** 1-13.

24. Raabe D, Mianroodi JR, Neugebauer J. Accelerating the design of compositionally complex materials via physics-informed artificial intelligence. *Nat Comput Sci* 2023, **3**(3)**:** 198-209.

25. Sasidhar KN, Siboni NH, Mianroodi JR, Rohwerder M, Neugebauer J, Raabe D. Enhancing corrosion-resistant alloy design through natural language processing and deep learning. *Sci Adv*, **9**(32)**:** eadg7992.

26. Ji Y, Li N, Cheng Z, Fu X, Ao M, Li M*, et al.* Random forest incorporating ab-initio calculations for corrosion rate prediction with small sample Al alloys data. *npj Mater Degrad* 2022, **6**(1)**:** 83.

27. Yong-fei J, Guo-shuai N, Yang Y, Yong-bing D, Jiao Z, Yan-feng H, *et al.* Knowledge-aware design of high-strength aviation aluminum alloys via machine learning. *J Mater Res Technol* 2023, **24:** 346-361.

28. Su H, Toda H, Shimizu K, Uesugi K, Takeuchi A, Watanabe Y. Assessment of hydrogen embrittlement via image-based techniques in Al–Zn–Mg–Cu aluminum alloys. *Acta Mater* 2019, **176:** 96-108.

29. Zhao H, Chakraborty P, Ponge D, Hickel T, Sun B, Wu C-H, *et al.* Hydrogen trapping and embrittlement in high-strength Al alloys. *Nature* 2022, **602**(7897)**:** 437-441.

30. Durodola J. Machine learning for design, phase transformation and mechanical properties of alloys. *Prog Mater Sci* 2022, **123:** 100797.

31. Ao M, Dong C, Li N, Wang L, Ji Y, Yue L, *et al.* Unexpected stress corrosion cracking improvement achieved by recrystallized layer in Al-Zn-Mg alloy. *J Mater Eng Perform* 2021, **30**(8)**:** 6258-6268.

32. Samek W, Binder A, Montavon G, Lapuschkin S, Müller K. Evaluating the visualization of what a deep neural network has learned. *IEEE Trans Neural Netw Learn Syst* 2017, **28**(11)**:** 2660-2673.

33. Wang F, Qiu D, Liu Z-L, Taylor JA, Easton MA, Zhang M-X. The grain refinement mechanism of cast aluminium by zirconium. *Acta Mater* 2013, **61**(15)**:** 5636-5645.

34. De Luca A, Dunand DC, Seidman DN. Microstructure and mechanical properties of a precipitation-strengthened Al-Zr-Sc-Er-Si alloy with a very small Sc content. *Acta Mater* 2018, **144:** 80-91.

35. Zhu Y, Sun K, Frankel G. Intermetallic phases in aluminum alloys and their roles in localized corrosion. *J Electrochem Soc* 2018, **165**(11)**:** C807.

36. Cheng S, Zhao Y, Zhu Y, Ma E. Optimizing the strength and ductility of fine structured 2024 Al alloy by nano-precipitation. *Acta Mater* 2007, **55**(17)**:** 5822-5832.

37. Li H, Yan Z, Cao L. Bake hardening behavior and precipitation kinetic of a novel Al-Mg-Si-Cu aluminum alloy for lightweight automotive body. *Mater Sci Eng, A* 2018, **728:** 88-94.

38. Chen H, Lu J, Kong Y, Li K, Yang T, Meingast A, *et al.* Atomic scale investigation of the crystal structure and interfaces of the B′ precipitate in Al-Mg-Si alloys. *Acta Mater* 2020, **185:** 193-203.

39. Lu L, Lai M, Hoe M. Formaton of nanocrystalline Mg2Si and Mg2Si dispersion strengthened Mg-Al alloy by mechanical alloying. *Nanostruct Mater* 1998, **10**(4)**:** 551-563.
40. Zandbergen MW, Xu Q, Cerezo A, Smith GDW. Study of precipitation in Al–Mg–Si alloys by Atom Probe Tomography I. Microstructural changes as a function of ageing temperature. *Acta Mater* 2015, **101:** 136-148.
41. Kairy SK, Birbilis N. Clarifying the role of Mg2Si and Si in localized corrosion of aluminum alloys by quasi in situ transmission electron microscopy. *Corrosion* 2020, **76**(5)**:** 464-475.
42. Ashtari P, Tezuka H, Sato T. Modification of Fe-containing intermetallic compounds by K addition to Fe-rich AA319 aluminum alloys. *Scripta Mater* 2005, **53**(8)**:** 937-942.
43. Li N, Dong C, Man C, Li X, Kong D, Ji Y, *et al.* Insight into the localized strain effect on micro-galvanic corrosion behavior in AA7075-T6 aluminum alloy. *Corros Sci* 2021, **180:** 109174.
44. Ji Y, Dong C, Kong D, Li X. Design materials based on simulation results of silicon induced segregation at AlSi10Mg interface fabricated by selective laser melting. *J Mater Sci Technol* 2020, **46:** 145-155.
45. Kharitonov DS, Dobryden I, Sefer B, Ryl J, Wrzesińska A, Makarova IV, *et al.* Surface and corrosion properties of AA6063-T5 aluminum alloy in molybdate-containing sodium chloride solutions. *Corros Sci* 2020, **171:** 108658.
46. Kairy SK, Rouxel B, Dumbre J, Lamb J, Langan TJ, Dorin T, *et al.* Simultaneous improvement in corrosion resistance and hardness of a model 2xxx series Al-Cu alloy with the microstructural variation caused by Sc and Zr additions. *Corros Sci* 2019, **158:** 108095.
47. Mantina M, Wang Y, Chen LQ, Liu ZK, Wolverton C. First principles impurity diffusion coefficients. *Acta Mater* 2009, **57**(14)**:** 4102-4108.
48. Peterson N, Rothman S. Impurity diffusion in aluminum. *Phys Rev B* 1970, **1**(8)**:** 3264.
49. Jiang L, Rouxel B, Langan T, Dorin T. Coupled segregation mechanisms of Sc, Zr and Mn at θ′ interfaces enhances the strength and thermal stability of Al-Cu alloys. *Acta Mater* 2021, **206:** 116634.
50. Gutiérrez RF, Sket F, Maire E, Wilde F, Boller E, Requena G. Effect of solution heat treatment on microstructure and damage accumulation in cast Al-Cu alloys. *J Alloys Compd* 2017, **697:** 341-352.

51. Norman AF, Prangnell PB, McEwen RS. The solidification behaviour of dilute aluminium–scandium alloys. *Acta Mater* 1998, **46**(16)**:** 5715-5732.

52. Marchand D, Curtin W. Machine learning for metallurgy IV: A neural network potential for Al-Cu-Mg and Al-Cu-Mg-Zn. *Phys Rev Mater* 2022, **6**(5)**:** 053803.

53. Sun Y, Peng Q, Lu G. Quantum mechanical modeling of hydrogen assisted cracking in aluminum. *Phys Rev B* 2013, **88**(10)**:** 104109.

54. Kresse G, Hafner J. Ab initio molecular dynamics for liquid metals. *Phys Rev B* 1993, **47**(1)**:** 558.

55. Perdew JP, Burke K, Ernzerhof M. Generalized gradient approximation made simple. *Phys Rev Lett* 1996, **77**(18)**:** 3865.

56. Hammer B, Hansen LB, Nørskov JK. Improved adsorption energetics within density-functional theory using revised Perdew-Burke-Ernzerhof functionals. *Phys Rev B* 1999, **59**(11)**:** 7413.

57. Gaillac R, Pullumbi P, Coudert F-X. ELATE: an open-source online application for analysis and visualization of elastic tensors. *J Phys: Condens Matter* 2016, **28**(27)**:** 275201.

58. Heyd J, Scuseria GE, Ernzerhof M. Hybrid functionals based on a screened Coulomb potential. *J Chem Phys* 2003, **118**(18)**:** 8207-8215.

59. Thompson AP, Aktulga HM, Berger R, Bolintineanu DS, Brown WM, Crozier PS*, et al.* LAMMPS-a flexible simulation tool for particle-based materials modeling at the atomic, meso, and continuum scales. *Comput Phys Commun* 2022, **271:** 108171.

60. Novikov IS, Gubaev K, Podryabinkin EV, Shapeev AV. The MLIP package: moment tensor potentials with MPI and active learning. *Mach Learn Sci Technol* 2021, **2**(2)**:** 025002.